\documentstyle[12pt,epsf]{article}
\def\lsim{<\kern-2.5ex\lower0.85ex\hbox{$\sim$}\ }
\def\rsim{>\kern-2.5ex\lower0.85ex\hbox{$\sim$}\ }
\input epsf

\def\ni{\noindent}
\def\LAMBDABAR
{\hbox{$\lambda$\kern-0.52em\raise+0.45ex\hbox{--}\kern+0.2em}}

\begin{document}

\baselineskip 18pt



\centerline{\large\bf{Measurement of Stochastic Gravitational Wave
Background}}

\centerline{\large\bf{with a Single Laser Interferometer}}

\vspace{.25in}

 \centerline{A.C. Melissinos and W.E. Butler}
 \centerline{\it Department of Physics and Astronomy, University of
 Rochester, Rochester, NY 14627}

\centerline{\today}

\vspace{.25in}

\begin{abstract}
Laser interferometer gravitational wave detectors can be operated
at their free spectral range frequency.  We show that in this case
and when the interferometer is well understood one could detect a
stochastic background using a single detector.
\end{abstract}

\ni {\bf 1. Introduction}

We define the stochastic gravitational wave background (sgwb) by a
dimensionless strain $h(t)$ which is a real random variable,
stationary in time.  Namely, it is Gaussian distributed with zero
mean
\begin{equation}
<h(t)>\ =\ 0
\end{equation}

\ni and finite \lq\lq power"
\begin{equation}
<h^2(t)>\ =\ P
\end{equation}

\ni Eqs.(1,2) do not suffice to completely describe the random
process.  To do so we must also specify how the power $P$ is
distributed in frequency\footnote{For instance whether the random
process is like \lq\lq white noise", \lq\lq 1/f noise", resonant
at a given frequency, etc.}, what is referred to as the power
spectral density (PSD).

We therefore assume that $h(t)$ can be represented, at a given
point in space, as a Fourier integral

\begin{equation}
h(t) = \int^\infty_{-\infty} \tilde{h}(f) e^{-2\pi i ft} df
\end{equation}

\ni Our assumptions about the properties of $h(t)$ imply that
\begin{equation}
<\tilde{h}(f)>\ = 0
\end{equation}

\ni and that [1]
\begin{equation}
<\tilde{h}^*(f)\tilde{h}(f^\prime)>\ = \frac{1}{2}
\delta(f-f^\prime)S(f)\qquad\qquad\qquad 0<f<\infty
\end{equation}

\ni The function $S(f)$ is the (one-sided) power spectral density
per unit time.  The proof of Eq.(5) and and the subtleties in the
definition of $\tilde{h}(f)$ are discussed in the Appendix.
Throughout this note we have adopted the nomenclature and
conventions of reference [2] as much as possible.

 The $\delta$-function can be removed from Eq.(5) by integrating
over $f^\prime$ in the narrow interval $f-\Delta f/2
<f^\prime<f+\Delta f/2$, to obtain
\begin{equation}
S(f) = ^{\lim}_{\Delta f\to 0} 2 \int^{f+\Delta f/2}_{f-\Delta
f/2} <\tilde{h}^*(f)\tilde{h}(f^\prime)> df^\prime \approx
2<|\tilde{h}(f)|^2> \Delta f
\end{equation}

\ni The integration over the small interval $\Delta f$ expresses a
kind of smoothing of $\tilde{h}(f)$ [3,4].  If $\tilde{h}(f)$ is
obtained from a stretch of data of length $T$, then $\Delta f =
1/T$ and Eq.(6) takes the form

\begin{equation}
S(f) = \frac{2}{T} <|\tilde{h}(f)|^2> \qquad\qquad\qquad
0<f<\infty
\end{equation}

\ni It follows from Parceval's theorem that

\begin{eqnarray}
\int^\infty_0 S(f)df & = & \frac{1}{T} \int^\infty_{-\infty} <
|\tilde{h}(f)|^2> \ df = \frac{1}{T} \int^{T/2}_{-T/2} |h(t)|^2dt
=\nonumber \\
\\
& = & < |h(t)|^2>\ = P\nonumber
\end{eqnarray}

 \ni Eqs.(7,8) are
our central result. Experimentally we wish to determine $S(f)$.
We will introduce the notation

\begin{equation}
\hat{h}(f) = \sqrt{S(f)}
\end{equation}

\ni and refer to $\hat{h}(f)$ as the sgwb amplitude (density) per
square root of frequency, i.e. $\hat{h}(f)$ represents strain
$/\sqrt{\rm Hz}$.

\vspace{.250in}

\ni{\bf 2. Single noiseless detector}

In an experiment we always deal with finite time intervals and
with time series (rather than functions of time).  This is well
suited to the form introduced in Eqs.(6,7).  We consider a linear
detector that in response to g.w. strain $h(t)$ outputs a signal
$a(t)$.  To include the response of the detector we must work in
the frequency domain

\begin{equation}
A(f) = H(f)\tilde{h}(f)
\end{equation}

\ni where $H(f)$ is a deterministic transfer function\footnote{We
take $H(f)$ to be dimensionless, that is the transfer function
from g.w.\,strain to strain recorded by a calibrated detector.}
peaked at some value $f_0$ and decreasing below and above $f_0$.
The detector output is given by

\begin{equation}
a(t) = \int^\infty_{-\infty} A(f) e^{-2\pi ift} df
\end{equation}

The output $a(t)$ is sampled at a rate $1/\Delta = 2f_c$, where
$f_c$ is referred to as the Nyquist critical frequency.  We
examine a record consisting of $N$ data points; the length of the
record is $\tau = N\Delta$ and the resulting time series is

\begin{equation}
a_n = a(t_n) \quad {\rm with}\quad t_n = n\Delta \qquad\qquad n =
0,1, \dots, (N-1)
\end{equation}

\ni We can form the Discrete Fourier Transform (DFT) of this time
series [2].

\begin{equation}
A_k = \sum^{N-1}_{n=0} a_n e^{2\pi ikn/N} \qquad\qquad\qquad\qquad
k = 0,1, \dots, (N-1)
\end{equation}

\ni This is a mapping of the $N$ complex numbers $a_n$ (of course
in our application the $a_n$ are real) onto $N$ complex numbers
$A_k$. The mapping is invertible

\begin{equation}
a_n = \frac{1}{N} \sum^{N-1}_{k=0} A_k e^{-2\pi ikn/N}
\end{equation}

\ni and obeys Parceval's theorem

\begin{equation}
\sum^{N-1}_{n=0} |a_n|^2 = \frac{1}{N} \sum^{N-1}_{k=0} |A_k|^2
\end{equation}

\ni The elements $A_k$ are related to $A(f_k)$, the Fourier
transform of $a(t)$, through

\begin{equation}
A(f_k) \approx \Delta A_k
\end{equation}

We can obtain an estimate of the PSD by forming the $periodogram$
of $a(t)$.  The frequency is defined only for zero and for $N/2$
positive values

\begin{equation}
f_k = \frac{k}{N\Delta} = 2f_c \frac{k}{N} \qquad\qquad\qquad k =
0,1, \dots,N/2
\end{equation}

\ni The bandwidth $B$ is the frequency spacing between $k$ and
$k+1$, or

\begin{equation}
B = 1/N\Delta = 1/\tau
\end{equation}

\ni The terms of the periodogram are

\begin{eqnarray}
P(0) & = & \frac{1}{N^2} |A_0|^2 \nonumber \\
\nonumber\\
P(f_k) & = & \frac{1}{N^2} \left[|A_k|^2 + |A_{-k}|^2\right]
\qquad\qquad\qquad k = 1,2,\dots,(N/2-1)\\
\nonumber\\
P(f_c) & =& \frac{1}{N^2} |A_{N/2}|^2\nonumber
\end{eqnarray}

\ni For $a(t)$ real $|A_k|^2 = |A_{-k}|^2$ and therefore

\begin{equation}
P(f_k) = \frac{2}{N^2} |A_k|^2
\end{equation}

We now use the approximate form of Eq.(16) in Eq.(7) to obtain for
the PSD {\it per unit time}

\begin{eqnarray}
S_A(f_k) & = & \frac{2}{\tau}<|A(f_k)|^2>\  \approx
\frac{2\Delta^2}{\tau} <|A_k|^2>\ = \nonumber \\
\\
& = & \tau \frac{2}{N^2}<|A_k|^2>\ = \frac{1}{B} P(f_k)\nonumber
\end{eqnarray}

\ni Returning to Eq.(10) we can write for the discrete case

\begin{equation}
\Delta A_k \approx A(f_k) =  H(f_k)\tilde{h}(f_k)
\end{equation}

\ni and therefore
\begin{equation}
S_A(f_k) = \frac{2}{\tau} |H(f_k)|^2 <|\tilde{h}(f_k)|^2>
\end{equation}

\ni If $\tilde{h}(f)$ is approximately constant in the region
where $H(f)$ peaks then the detected spectral density (per unit
time) $S_A(f_k)$ {\it will follow the spectrum of} $|H(f_k)|^2$.

\vspace{.25in}

 \ni{\bf 3. Single noisy detector}

Suppose now that the detector introduces noise so that the output
signal is

\begin{equation}
c(t) = q(t) + a(t)
\end{equation}

\ni Here $q(t)$ is a stochastic variable with zero mean and
standard deviation $\sigma$.  After sampling $c(t)$ and forming
the DFT we obtain as in Eq.(13).

\begin{equation}
C_k = Q_k + A_k
\end{equation}

\ni Since $q(t)$ and $a(t)$ are uncorrelated, the PSD/time, upon
averaging, will contain only two terms

\begin{equation}
S_C(f_k) = \frac{2\tau}{N^2} \left[<|Q_k|^2> + <|A_k|^2>
\frac{}{}\right]
\end{equation}

If the noise spectrum is flat in frequency, one can extract
$<|A_k|^2>$ by fitting $S_C(f_k)$ with the expected spectral
shape, which according to Eq.(23) is that of $|H(f_k)|^2$.  In
this way one can determine $<|\tilde{h}(f_k)|^2>$, in the region
$f_k=f_0$, with a single detector, provided the spectrum of
$<|Q_k|^2>$ is known.

In fact, one can work directly with the magnitude of $A_k$ which
is related to $\hat{h}(f)$, defined by Eq.(9).  To extract a
meaningful signal, $|A_k|$ must be larger or equal than the
$fluctuations$ of $Q_k$ (rather than $Q_k$ itself).  For a single
measurement the fluctuations of $|Q_k|$ are equal to
$\left[<|Q_k|^2>\frac{}{}\right]^{1/2}$ since $q(t)$ was assumed
to have zero mean.  After averaging $N_a$ different DFT spectra

\begin{equation}
\sigma_{Q_k}(N_a) = \frac{1}{\sqrt{N_a}} |Q_k|
\end{equation}

\ni Further, in fitting the data to extract $|A_k|^2$ we have at
our disposal the $n$ frequency bins in the bandwidth $\Delta f$,
typically a few times the FWHM of $H(f)$,

\begin{equation}
n = \Delta f/B = \tau\Delta f
\end{equation}

The significance of the fit can be expressed as a signal to noise
ratio

\begin{eqnarray}
\frac{S}{N}|_{N_a} & = & \frac{\sqrt{<|A_k|^2>}}{\sqrt{<|Q_k|^2>}}
\sqrt{n} \sqrt{N_a}
= \sqrt{T\Delta f} \frac{\sqrt{<|A_k|^2>}}{\sqrt{<|Q_k|^2>}} =\nonumber \\
\\
& = & \sqrt{T\Delta f} \overline{|H_k|}
\left[\frac{\hat{h}_s}{\hat{h}_N}\right]\nonumber
\end{eqnarray}

\ni In Eq.(29) we have set $T = \tau N_a$ for the total
measurement time and $\overline{|H_k|}$ is the magnitude of the
transfer function averaged over the frequency band $\Delta f$.  We
also used Eqs.(25,26) and the definition of Eq.(9) to relate
$<|A_k|>$ and $<|Q_k|>$ to the amplitude (densities) per unit
frequency for the sgwb signal and the noise in the detector.

 Eq.(29) shows that when using a single interferometer, the $S/N$ ratio for
$\left[\hat{h}_s/\hat{h}_N\frac{}{}\right]$ improves as
$\sqrt{T\Delta f}$, in contrast to the correlation method [see
Eq.(37)] where the $S/N$ for the same ratio improves only as
$(T\Delta f)^{1/4}$.  This is particularly true if no signal is
detected and the measurement yields only an upper limit on
$\hat{h}_s$. Computer simulations [5] of a sgwb signal using for
noise the actual time series from LIGO test data confirm this
conclusion.

\vspace{.25in}

\ni{\bf 4. Correlated detectors}

The \lq\lq standard" approach to the measurement of a stochastic
signal is to correlate the output of two detectors.  For the sgwb
this method has been discussed extensively in [6,7] and has been
applied to data from the LIGO interferometers in [8].  Here we
only briefly outline the analysis procedure and present the
conclusions.

Assuming that the two detectors 1,2 are co-located and co-aligned,
the stochastic signal, $a(t)$ will be the same in both detectors
whereas the noise signals $q_1(t)$ and $q_2(t)$ are treated as
uncorrelated.  Thus we write

\begin{eqnarray}
c_1(t) & = & q_1(t) + a(t)\nonumber \\
\\
c_2(t) & = & q_2(t) + a(t)\nonumber
\end{eqnarray}

We form the cross-correlation (at zero lag) of the two signals and
integrate over a time interval $\tau$, to obtain a statistic [1]

\begin{equation}
C = \int^\tau_0 c_1(t) c_2(t) dt
\end{equation}

\ni The mean value $\mu$, of this statistic obtained from repeated
cross-correlation spectra (samples) is a measure of the stochastic
signal because

\begin{equation}
\mu = \ <C>\ = \int^\tau_0 <a^2(t)>dt = \tau<a^2(t)>
\end{equation}

\ni We can express $\mu$ in terms of frequency domain variables by
using the definition of Eq.(8)

\begin{equation}
\mu = \tau \int^\infty_0 S_A(f) df
\end{equation}

\ni where $S_A(f)$ is the spectral density of the stochastic
signal.

 The fluctuations in $\mu$ are given by the variance of $C$

\begin{equation}
\sigma^2_C =\ <C^2> - <C>^2
\end{equation}

\ni and if $a(t) \ll q_i(t),\ \sigma_C^2$ is dominated by the
noise spectrum

\begin{equation}
\sigma_C^2 = \int^\tau_0 <q_1^2(t)> <q_2^2(t)> d\tau = \tau
\int^\infty_{0} S_{N_1}(f)S_{N_2}(f)df
\end{equation}

\ni where $S_{Ni}(f)$ are the  spectral densities of the noise. If
we use $N_S$ samples to determine the mean, then $<\mu>$ will be
Gaussian distributed around its true value with standard deviation
$\sigma_C/\sqrt{N_S}$. Thus the signal to noise ratio is

\begin{equation}
\frac{S}{N}|_{N_S} = \frac{<\mu>}{\sigma_C} \sqrt{N_S} =
\sqrt{N_S\tau}\ \ \frac{\int^\infty_0 S_A(f)df}{[\int^\infty_0
S_{N_1}(f)S_{N_2}(f)df]^{1/2}}
\end{equation}

If we further assume that $S_A(f)$ is reasonably constant in the
region of integration, and so are $S_{N_1}(f) \simeq S_{N_2}(f)$
we can use the notation of Eq.(9) to simplify Eq.(36)

\begin{equation}
\frac{S}{N}|_{N_S} = \sqrt{T\Delta f}
\left[\frac{\hat{h}_s(f)}{\hat{h}_N(f)}\right]^2
\end{equation}

\ni where we set $T = N_S \tau$ for the total measurement time.
Eq.(37) is to be compared with Eq.(29) obtained for measurements
with a single detector.

\vspace{.25in}

 \ni {\bf Acknowledgements}

This note has been motivated by the work of ref. [5] where an
analysis along the lines proposed here was carried out. We are
grateful to E. Wolf for critical discussions and for reading an
early draft of this note, and to M. Bocko for clarifying
suggestions. We also thank N. Stull for computer simulations, and
S. Giampanis for constructive comments.

\vfil\eject

\centerline{\bf APPENDIX A}

\vspace{.25in}
 \centerline{\bf Spectral Density of a Random Variable}

\vspace{.25in}

 \ni We expressed a stochastic signal $h(t)$ by Eq.(3)

$$
h(t) = \int^\infty_{-\infty} \tilde{h}(f) e^{-2\pi ift} df
\eqno{(\rm A1)}
$$

\ni It is not possible to invert Eq.(A1) because the integral

$$\int^\infty_{-\infty} h(t) e^{2\pi ift} dt \eqno{(\rm A2)}$$

\ni diverges for a stationary signal.

Instead, following [4] we can introduce the truncated function

$$
h_T(t) = h(t) \qquad\qquad\qquad -T/2<t<T/2$$
$$\quad \ \ =  0 \qquad\qquad\qquad\qquad {\rm elsewhere}\qquad \eqno{\rm (A3)}
$$

\ni Thus

$$\tilde{h}_T(f) =\int ^\infty_{-\infty} h_T(t) e^{2\pi ift} dt
\eqno {(\rm A4)}$$

\ni and we form

$$\tilde{h}^*_T(f) \tilde{h}_T(f^\prime) =
\int^\infty_{-\infty}\int^\infty_{-\infty} h^*_T(t)
h_T(t^\prime)e^{-2\pi ift} e^{2\pi if^\prime t^\prime} \ dt
dt^\prime \eqno{(\rm A5)}$$

\ni We set $t^\prime = t + \tau$ so that

$$\tilde{h}^*_T(f) \tilde{h}_T(f^\prime) = \int^\infty_{-\infty}
\int^\infty_{-\infty} h^*_T(t) h_T (t + \tau) e^{2\pi
i(f^\prime-f)t} e^{2\pi if^\prime \tau} dtd\tau \eqno{(\rm A6)}$$

\ni We now take the ensemble average of both sides, interchange
the order of averaging and integration on the right-hand side, and
define the {\it autocorrelation} function [9, 3]

$$R_T(\tau) =\  <h^*_T(t) h_T(t+\tau)> \eqno{(\rm A7)}$$

\ni The function $R_T(\tau)$ vanishes for $|\tau|>T$, and Eq. (A6)
reads

$$<\tilde{h}^*_T(f) \tilde{h}_T(f^\prime)>\ = \delta(f-f^\prime)
\int^\infty_{-\infty} R_T(\tau) e^{2\pi if\tau} d\tau \eqno{(\rm
A8)}$$

\ni We therefore define the power spectral density (per unit time)
as the Fourier transform of the autocorrelation function

$$S_T(f) = 2\int^\infty_{-\infty} R_T(\tau) e^{2\pi if \tau} d\tau
\qquad\qquad\qquad 0<f<\infty \eqno{(\rm A9)}$$

\ni Thus Eq.(A8) is recast as

$$<\tilde{h}^*_T (f) \tilde{h}_T (f^\prime)>\ = \frac{1}{2} \delta
(f-f^\prime) S_T(f)\eqno{(\rm A10)}$$

\ni By letting $T \to \infty$ we obtain Eq.(5) of the main text.

To prove Eq.(7) of the main text we proceed as follows.  By
definition, [Eq.(A7)],

$$
R_T (0) = \ <|h_T(t)|^2>\  =  \frac{1}{T} \int^{T/2}_{-T/2}
|h_T(t)|^2dt = \frac{1}{T} \int^\infty_{-\infty} |h_T(t)|^2dt =
$$
$$
 =  \frac{1}{T} \int^\infty_{-\infty}
|\tilde{h}_T(f)|^2df\eqno{(\rm A11)}
$$

\ni Integrating Eq.(A9) over frequency we obtain

$$\int^\infty_0 S_T(f)df = \frac{1}{2}\int^\infty_{-\infty}
2R_T(\tau) e^{2\pi if\tau} dfd\tau = \int^\infty_{-\infty}
R_T(\tau)\delta(\tau) d\tau = R_T(0) =$$

$$ =
\frac{1}{T}\int^\infty_{-\infty}|\tilde{h}_T(f)|^2df = \frac{2}{T}
\int^\infty_0 |\tilde{h}_T(f)|^2df \qquad\eqno{(\rm A12)}$$

\ni where we used Eq.(A11) for $R_T(0)$.  Equating the integrands
we establish

$$S_T(f) = \frac{2}{T} |\tilde{h}_T(f)|^2\eqno{(\rm A13)}$$

\ni which is analogous to Eq.(7) of the text but for the Fourier
transforms $\tilde{h}_T(f)$ of the {\it truncated functions}
$h_T(t)$.  It would seem that by letting $T\to \infty$ we could
recover the relation shown in Eq.(7).  This is not the case [10]
because $S_T(f)$ does not converge to a single value as $T \to
\infty$.  It can however be \lq\lq smoothed" by taking the
ensemble average of $|h_T(f)|^2$ {\it before} letting $T \to
\infty$, so that

$$S(f) = ^{\lim}_{T\to \infty} \frac{2}{T} <|\tilde{h}_T(f)|^2>\eqno{(\rm A14)}$$

\ni which is the mathematically precise statement of Eq.(7).

\vspace{.50in}

It is of some interest to express Eq.(A10) in the time domain.
Using the inverse of Eq.(A4) we write

$$h^*_T(t)h_T(t^\prime) =
\int^\infty_{-\infty}\int^\infty_{-\infty} \tilde{h}^*(f) e^{2\pi
ift} \tilde{h}(f^\prime) e^{-2\pi if^\prime t^\prime} dfdf^\prime
\eqno{(\rm A15)}$$

\ni Ensemble average, interchange the order of integration and
averaging, and use Eq.(A10)

$$<h^*_T(t) h_T(t^\prime)>\ = \frac{1}{2}\int^\infty_{-\infty}
\int^\infty_{-\infty} \delta (f-f^\prime) S_T(f) e^{2\pi
i(ft-f^\prime t^\prime)} df^\prime df$$
$$ = \frac{1}{2}\int^\infty_{-\infty} e^{2\pi if(t-t^\prime)}
S_T(f) df \eqno {(\rm A16)}$$

\ni When $t' = t$, Eq.(A16) reduces to Eq.(8) of the main text in
the limit $T\to \infty$.

\vspace{.50in}

The physical interpretation of the autocorrelation function,
Eq.(A7), implies that if the signal $h(t)$ \lq\lq repeats" itself,
on the average, with period $\tau_0$, then it must contain a
frequency component $f=1/\tau_0$.  For example if $h(t) =
\cos\omega_0t = \cos(2\pi f_0t) $, then

$$R_T(\tau) = \cos(2\pi f_0\tau) \qquad {\rm and} \qquad S_T(f) =
\delta (f-f_0) \eqno{(\rm A17)}$$

\ni If $R_T(\tau)$ drops off for $\tau > \tau_0$, then the highest
frequency contained in $h(t)$ is $f<1/\tau_0$.  For white noise we
can write $R_T(\tau) = \tau_c\delta(\tau)$  where $\tau_c$ is the
correlation time, defined for instance by our sampling interval.
Then the spectral density is flat up to a frequency $f_c =
1/\tau_c$.

\vfil\eject

\ni {\bf References and Notes}

\begin{enumerate}

\item B. Allen and J. Romano, Phys. Rev. {\bf D59} (1999), 102001.

\item W.H. Press, S.A. Teukolsky, W.T. Vetterling and B.P.
Flannery, {\it Numerical Recipes}, Cambridge (1992).

\item L. Mandel and E. Wolf, {\it Optical Coherence and Quantum
Optics}, Cambridge (1995).

\item S. Goldman, {\it Information Theory}, Prentice Hall (1995).
See in particular section 8.4.

 \item W.E. Butler, {\it
Characterization of the High Frequency Response of Laser
Interferometer Gravitational Detectors:}, Thesis, University of
Rochester 2004, \linebreak http://www.ligo.caltech.edu/document
number P040026-00-R.

\item P.F. Michelson, Mon. Not. R. Astron. Soc. {\bf 227}, 922
(1987).

\item E.E. Flanagan,  Phys. Rev. {\bf D48}, 2389 (1993).

\item B. Abbott et al., Phys. Rev. {\bf D69}, 122004 (2004).

\item N. Wiener, Acta Math. {\bf 55}, 117-258 (1930).

\item E. Wolf, J. Opt. Soc. Am. {\bf 72}, 343 (1982).

\end{enumerate}

\end{document}